\documentclass[12pt]{article}
\usepackage{epsfig,amsfonts,amsthm}
\usepackage{amsmath,amssymb}
\usepackage{cite}
\usepackage{xcolor}
\usepackage{tikz}

\newcommand{\be}{\begin{equation}}
\newcommand{\ee}{\end{equation}}
\newcommand{\bea}{\begin{eqnarray}}
\newcommand{\eea}{\end{eqnarray}}

\newcommand{\triplet}[3]{ \left(\! \begin{array}{c}#1 \\ #2 \\ #3 \end{array}\!\right) }
\providecommand{\mtrx}[1]{\begin{pmatrix} #1 \end{pmatrix}}
\newcommand{\lr}[1]{ \langle #1 \rangle}

\newcommand{\Z}{\mathbb{Z}}
\newcommand{\mmmatrix}[9]{ \left(\!\! \begin{array}{ccc}#1 & #2 & #3\\ #4 & #5 & #6\\ #7 & #8 & #9\\ \end{array}\!\!\right) }
\providecommand{\id}{{\boldsymbol{1}}}

\def\lsim{\mathrel{\rlap{\lower4pt\hbox{\hskip1pt$\sim$}}
    \raise1pt\hbox{$<$}}}         
\def\gsim{\mathrel{\rlap{\lower4pt\hbox{\hskip1pt$\sim$}}
    \raise1pt\hbox{$>$}}}         

\title{Exploring multi-Higgs models with softly broken large discrete symmetry groups}

\author{Ivo~de~Medeiros~Varzielas$^{1}$\thanks{E-mail: ivo.de@udo.edu},	
	Igor~P.~Ivanov$^{2}$\thanks{E-mail: ivanov@mail.sysu.edu.cn}, 
	Miguel Levy$^{1}$\thanks{E-mail: miguelplevy@ist.utl.pt}   
	\\
	{\small $^1$ CFTP, 
		Instituto Superior T\'{e}cnico, Universidade de Lisboa, 
		av. Rovisco Pais 1, 1049-001 Lisboa, Portugal}\\
	{\small $^2$ School of Physics and Astronomy, Sun Yat-sen University, 519082 Zhuhai, China}
}

\begin{document}
\maketitle

	\bigskip
\begin{abstract}
We develop methods to study the scalar sector of multi-Higgs models with large discrete symmetry groups that are softly broken. 
While in the exact symmetry limit, the model has very few parameters and can be studied analytically, 
proliferation of quadratic couplings in the most general softly broken case makes the analysis cumbersome.
We identify two sets of soft breaking terms which play different roles: 
those which preserve the symmetric vacuum expectation value alignment,
and the remaining terms which shift it. 
Focusing on alignment preserving terms, we check which structural features of the symmetric parent model
are conserved and which are modified.
We find remarkable examples of structural features which are inherited from the parent symmetric model
and which persist even when no exact symmetry is left.
The general procedure is illustrated with the example of the three-Higgs-doublet model 
with the softly broken symmetry group $\Sigma(36)$.
\end{abstract}


\section{Introduction}

\subsection{Taming the large number of free parameters}

Numerous pieces of evidence suggest that the Standard Model (SM) cannot be the ultimate theoretical
construction of the microscopic world. In the absence of direct compelling hints
of how New Physics beyond the SM should look like, theorists explore different venues.
A very active direction of research 
is the study of non-minimal scalar sectors,
for a selection of topics see the recent reviews \cite{Branco:2011iw,Kanemura:2014bqa,Ivanov:2017dad,Arcadi:2019lka}.
The simple idea that Higgs doublets can come in generations, just like fermions, 
alleviates some of the problems of the SM and also leads to a surprisingly rich list of phenomena.
After a decades-long study of the two-Higgs-doublet models (2HDMs) \cite{Branco:2011iw}, 
the community is exploring other scalar sectors, such as the three-Higgs-doublet models (3HDMs).

First proposed by S.~Weinberg in 1976 \cite{Weinberg:1976hu}, the 3HDMs equipped with various global symmetries,
exact or softly broken, were studied in hundreds of papers, see a brief historical overview in \cite{Ivanov:2017dad}.
However, a systematic study of all the opportunities offered by the 3HDMs is still lacking.
One obvious reason for that is the very large number of free parameters.
The most general renormalizable scalar potential of the $N$-Higgs-doublet model can be written, at the tree level, as
\begin{equation}
V=Y_{ij} (\phi^{\dagger}_i\phi_j)+Z_{ijkl} (\phi^{\dagger}_i\phi_j)(\phi^{\dagger}_k\phi_l)\,, \quad i,j,k,l = 1, \dots, N,\label{YZ}
\end{equation}
with 54 free parameters for $N=3$. If one includes the quark Yukawa sector, the total number of free parameters
exceeds one hundred. Certainly, it is possible, for any particular point in the entire parameter space, 
to numerically minimize the potential, compute all scalar masses and couplings, track down the fermion sector 
and its interaction with new scalars.
But the real challenge is to make sense of these case-by-case calculations and to identify 
all the essentially distinct phenomenological situations which may be hiding in various parts of the very-large-dimensional 
parameter space.
The richness of the 3HDM is just too vast to grasp and visualize with a straightforward (numerical) approach. 

One popular way to tame the proliferation of free parameters is to assume that the multi-Higgs model
is equipped with an additional global symmetry group. In early 1980's, this approach seemed promising because one hoped to link the mixing angles of the 
Cabibbo-Kobayashi-Maskawa matrix with quark mass ratios \cite{Ivanov:2017dad}.
Later it turned out that this direct approach exploiting the exact symmetry groups could not lead to a viable quark sector
\cite{Leurer:1992wg,GonzalezFelipe:2014mcf},
but softly broken symmetries seemed to offer sufficient flexibility and interesting predictions.
In the case of the 3HDM scalar sector, several continuous and discrete symmetry groups have been implemented,
starting from Weinberg's model, which has the symmetry group $\Z_2\times \Z_2$.
The full classification of discrete symmetry groups usable in the scalar sector of the 3HDM was established in \cite{Ivanov:2012fp}.
If one focuses on the Higgs potential alone, one can observe accidental symmetries which go beyond Higgs family transformations.
They were classified in \cite{Darvishi:2019dbh} and a deeper analysis of the so-called maximally symmetric 3HDM
was presented in \cite{Darvishi:2021txa}.
The full list of symmetry groups suitable for the 3HDM Yukawa sector is still not known. 
The $CP$ properties of the 3HDM scalar sector were also explored by using basis-independent methods \cite{Nishi:2006tg, deMedeirosVarzielas:2017ote, Ivanov:2018ime, Ivanov:2019kyh, deMedeirosVarzielas:2019rrp}.

\subsection{Softly broken large discrete symmetry}

Investigation of the 3HDMs with softly broken global symmetry group $G$ depends on the group itself.
Let us focus on the attractive case of large discrete groups $G$, with Higgs doublets transforming
as an irreducible triplet representations. Four such cases are known\footnote{The 3HDM with a $\Delta(27)$ triplet leads to the same potential as the potential for a $\Delta(54)$ group, so it is implicitly included in the $\Delta(54)$ case.} \cite{Ivanov:2012fp}:
$G = A_4$, $S_4$, $\Delta(54)$, and $\Sigma(36)$-symmetric 3HDMs. In any of these four cases,
the Higgs potential invariant under $G$ has the following form:
\begin{equation}
V_0 = - m^2 (\phi_1^\dagger \phi_1 + \phi_2^\dagger \phi_2 + \phi_3^\dagger \phi_3) + V_4\,,
\end{equation}
while the $G$-symmetric quartic potential $V_4$ contains several terms.
All possible minima of this potential for the symmetry groups $A_4$, $S_4$, $\Delta(54)$, and $\Sigma(36)$
are known and were put together in \cite{Ivanov:2014doa} (see also \cite{deMedeirosVarzielas:2017glw}).

If one wants to explore a 3HDM with a softly broken group $G$, one needs to introduce all possible
quadratic terms:
\begin{eqnarray}
V_{\rm soft} = m_{11}^2 \phi_1^\dagger\phi_1 + m_{22}^2 \phi_2^\dagger\phi_2 + m_{33}^2 \phi_3^\dagger\phi_3 
+ \left(m_{12}^2\,\phi_1^\dagger\phi_2 + m_{23}^2\,\phi_2^\dagger\phi_3 
+ m_{31}^2\,\phi_3^\dagger\phi_1 + h.c.\right)\label{soft}
\end{eqnarray}
with complex $m_{ij}^2$ for $i \not = j$.
In total, there are nine free parameters here. Exploring in detail the emerging phenomenology in all corners of this 9-dimensional
soft breaking parameter space and visualizing the results would be very challenging.
However these free parameters do not play equal roles.
Some parameters may trigger structural changes,
while others will only shift the numerical values of the observables.
Some phenomena may happen along generic directions in this soft breaking parameter space, 
while other effects may take place only along some very particular directions.
One could even think of plotting a phase diagram of the phenomenology of the resulting 3HDM with softly broken $G$,
but describing it in its full dimensionality seems very hard.

In short, one needs a guiding principle and a set of efficient methods to make sense of multi-Higgs
models with softly broken large discrete symmetry groups.

This is the main goal of the present paper.
We will show that the nine soft breaking free parameters can be split into two families:
five parameters which preserve the vacuum expectation value (vev) alignment of the exactly symmetric parent model,
and the four parameters which drive this alignment away in four orthogonal directions.
Focusing on the vev-preserving parameters, we will show which structural features of the fully symmetric model
stay unchanged and which get modified, and how to track the effect of each of these parameters.
In particular, we will find that, although the models with vev-preserving soft breaking terms do not possess any exact symmetry,
their scalar sector phenomenology ``inherits'' some of the features from the parent $G$-symmetric model.
These results help develop qualitative and quantitative intuition when building multi-Higgs models 
with softly broken large discrete symmetry groups.

These phenomena will be illustrated with the example of the largest discrete symmetry group possible
in the 3HDM scalar sector, the group $\Sigma(36)$, which is not as well known as the $A_4$, $S_4$, or $\Delta(54)$-symmetric 3HDMs.
The application of the methodology developed here to those groups and the inclusion of the Yukawa sector are delegated to future works.

The structure of the paper is the following.
In the next section we outline the properties of the scalar sector emerging in the $\Sigma(36)$-symmetric 3HDM.
In section~\ref{section-soft} we construct the soft breaking terms which preserve any chosen vev alignment
of the symmetric model and illustrate the general procedure with the softly broken $\Sigma(36)$ 3HDM.
Finally, we discuss the results and draw conclusions, while the Appendix provide auxiliary mathematical details.

\section{$\Sigma(36)$-symmetric 3HDM}\label{section-exact}

\subsection{The scalar potential and its minima}

$\Sigma(36)$ is the largest discrete symmetry group which can be imposed on the scalar sector of the 3HDM
without leading to accidental continuous symmetries \cite{Ivanov:2012fp}.
Group-theoretically, it is defined as a $\Z_4$ permutation acting on generators of the abelian group 
$\Z_3\times \Z_3$:
\begin{equation}
\Sigma(36) \simeq (\Z_3 \times \Z_3)\rtimes \Z_4\,.
\end{equation}
The generators of the $\Z_3\times \Z_3$ ``core'' and the generator of $\Z_4$ are
\begin{equation}
a = \mtrx{1&0&0\\ 0&\omega&0\\ 0&0&\omega^2}\,, \quad
b = \mtrx{0&1&0\\ 0&0&1\\ 1&0&0}\,,\quad
d = {i \over\sqrt{3}} \left(\begin{array}{ccc} 1 & 1 & 1 \\ 1 & \omega^2 & \omega \\ 1 & \omega & \omega^2 \end{array}\right)\,,
\label{Sigma36-generators}
\end{equation}
where $\omega = \exp(2\pi i/3)$. The orders of these generators are:
\[
a^3 = \id\,,\quad b^3 = \id\,, \quad d^4 = \id\,.
\]
Notice that $d^2$ is a transformation which transposes two doublets; thus, $\Sigma(36)$ contains all permutations of the three doublets.
Had we imposed symmetry under $d^2$ but not $d$, 
we would end up with the more familiar symmetry group $\Delta(54)$,
first used within the 3HDMs back in late 1970's \cite{Segre:1978ji} and explored later in \cite{Grimus:2010ak, Merle:2011vy,Hagedorn:2013nra, Rong:2016cpk}.
For more details on the relation between $\Delta(54)$ and $\Sigma(36)$ and the subtleties of their definitions, see Appendix~\ref{appendix-Delta54}.

The scalar potential of 3HDM invariant under $\Sigma(36)$ has the following form:
\begin{eqnarray}
V_0 & = &  - m^2 \left[\phi_1^\dagger \phi_1+ \phi_2^\dagger \phi_2+\phi_3^\dagger \phi_3\right]
+ \lambda_1 \left[\phi_1^\dagger \phi_1+ \phi_2^\dagger \phi_2+\phi_3^\dagger \phi_3\right]^2 \nonumber\\
&&
- \lambda_2 \left[|\phi_1^\dagger \phi_2|^2 + |\phi_2^\dagger \phi_3|^2 + |\phi_3^\dagger \phi_1|^2
- (\phi_1^\dagger \phi_1)(\phi_2^\dagger \phi_2) - (\phi_2^\dagger \phi_2)(\phi_3^\dagger \phi_3)
- (\phi_3^\dagger \phi_3)(\phi_1^\dagger \phi_1)\right] \nonumber\\
&&
+ \lambda_3 \left(
|\phi_1^\dagger \phi_2 - \phi_2^\dagger \phi_3|^2 +
|\phi_2^\dagger \phi_3 - \phi_3^\dagger \phi_1|^2 +
|\phi_3^\dagger \phi_1 - \phi_1^\dagger \phi_2	|^2\right)\, .
\label{Vexact}
\end{eqnarray}
It has four real free parameters. The first two lines of Eq.~\eqref{Vexact} are invariant under the entire $SU(3)$
transformation group of the three Higgs doublets.
The positive sign of $\lambda_2$ guarantees that the minimum corresponds to a neutral vacuum,
but the minimization of these two lines alone would lead to several neutral Nambu-Goldstone bosons.
The last term with the coefficient $\lambda_3$ selects the discrete $\Sigma(36)$ group out of it
and renders those Higgs bosons massive.

The potential of Eq.~\eqref{Vexact} is also $CP$ invariant. Apart from the standard $CP$ symmetry $\phi_i \mapsto \phi_i^*$,
it is also invariant under many other $CP$ transformations of the form of the standard $CP$ combined with any of the 
symmetries from $\Sigma(36)$. Unlike the $\Delta(54)$ 3HDM, the absence of $CP$ violation in $\Sigma(36)$ 3HDM is not an assumption
but is a consequence of the $\Z_4$ subgroup
which forbids any form of $CP$ violation in 3HDM, explicit or spontaneous \cite{Ivanov:2011ae,Ivanov:2014doa}.

An important feature of the entire $\Delta(54)$ family of 3HDM models, including $\Sigma(36)$ 3HDM, is the rigid structure of its minima.
Depending on the values of the parameters, the global minimum of the potential can only correspond to the following vev alignments \cite{Ivanov:2014doa}:
\begin{eqnarray}
\mbox{alignment $A$:}&& A_1 = (\omega,\,1,\,1)\,, \quad A_2 = (1,\,\omega,\,1),\, \quad A_3 = (1,\,1,\,\omega)\label{points-A}\\
\mbox{alignment $A'$:}&& A'_1 = (\omega^2,\,1,\,1)\,, \quad A'_2 = (1,\,\omega^2,\,1),\, \quad A'_3 = (1,\,1,\,\omega^2)\label{points-Ap}\\
\mbox{alignment $B$:}&& B_1 = (1,\,0,\,0)\,, \quad B_2 = (0,\,1,\,0),\, \quad B_3 = (0,\,0,\,1)\label{points-B}\\
\mbox{alignment $C$:}&& C_1 = (1,\,1,\,1)\,,\quad C_2 = (1,\,\omega,\,\omega^2)\,,\quad C_3 = (1,\,\omega^2,\,\omega)\label{points-C}
\end{eqnarray}
Notice that, due to the global symmetry of the 3HDM potential under simultaneous phase rotation of the three doublets,
other possible configurations can be reduced to these ones; for example, $(\omega,\,\omega^2,\, 1) = \omega (1,\,\omega,\,\omega^2)$
also corresponds to the alignment $C_2$.
The phase rigidity is reflected in the fact that the relative phases between vevs are calculable and are not sensitive
to the exact numerical values of the coefficients.
This rigidity was behind the ``geometric $CP$-violation'' proposal back in 1984 \cite{Branco:1983tn} which was revisited in more detail in \cite{deMedeirosVarzielas:2011zw, Varzielas:2012nn, Ivanov:2013nla}, and also shown to be compatible with viable Yukawa sectors \cite{Bhattacharyya:2012pi, Varzielas:2013sla, Varzielas:2013eta}.

None of the minima of the $\Sigma(36)$-symmetric 3HDM breaks the symmetry group completely~\cite{Ivanov:2014doa}.
There are six family symmetries and six $CP$-type symmetries which are preserved at each vev alignment.
Other, spontaneously broken, symmetries link different vacua, which, despite looking differently, represent the same physics.
For $\lambda_3 < 0$, the global minima are at the six points $A$ and $A'$, which are related 
by the broken symmetries from $\Sigma(36)$.
For $\lambda_3 > 0$, the degenerate global minima are at points $B$ or $C$. 
Thus, we have two essentially distinct phenomenological situations in $\Sigma(36)$ 3HDM.

Further insights into the structural properties of the model, including the vev alignments, symmetry and $CP$ properties, 
can be gained if one pays attention not only to the transformations from the symmetry group $G$
but also to the transformations from $SU(3)$ which leave $G$ invariant, 
or ``symmetries of symmetries'' in the language of \cite{Fallbacher:2015rea}.
The potential remains form-invariant under such transformations, only up to reparametrizatioon of coefficients,
which may provide additional links between different regimes of the same model.

\subsection{The physical Higgs bosons}

Three Higgs doublets contain 12 real fields. When expanding the potential around 
a neutral vacuum, one absorbs, as usual, three of them in the longitudinal components of
the $W^\pm$ and $Z$-bosons. What remains is two pairs of charged Higgses and five neutral Higgs bosons.
At points $B$ or $C$, the Higgs boson masses are
\begin{eqnarray}
m_{h_{SM}}^2 &=& 2\lambda_1 v^2 = 2m^2\,, \nonumber\\[1mm]
m_{H^\pm}^2 &=& {1\over 2}\lambda_2 v^2\quad \mbox{(double degenerate)}\,,\nonumber\\[1mm]
m_{h}^2 &=& {1\over 2}\lambda_3 v^2\quad \mbox{(double degenerate)}\,,\nonumber\\[1mm]
m_{H}^2 &=& 3 m_h^2 = {3\over 2}\lambda_3 v^2\quad \mbox{(double degenerate)}\,.\label{masses-BC}
\end{eqnarray}
At points $A$ and $A'$, which are the minima for $\lambda_3 < 0$, the Higgs masses are 
\begin{eqnarray}
m_{h_{SM}}^2 &=& 2(\lambda_1 +\lambda_3)v^2 = 2m^2\,, \nonumber\\[1mm]
m_{H^\pm}^2 &=& {1\over 2}(\lambda_2 - 3\lambda_3)v^2\quad \mbox{(double degenerate)}\,,\nonumber\\[1mm]
m_{h}^2 &=& -{1\over 2}\lambda_3 v^2\quad \mbox{(double degenerate)}\,,\nonumber\\[1mm]
m_{H}^2 &=& 3 m_h^2 = -{3\over 2}\lambda_3 v^2\quad \mbox{(double degenerate)}\,.\label{masses-AA}
\end{eqnarray}
Identification of the SM-like Higgs boson is unambiguous. It is a straightforward exercise to show
that, if the quadratic potential of an NHDM has the form $m^2\sum_i \phi_i^\dagger\phi_i$,
then whatever the quartic potential is, the model automatically incorporates the exact {\em scalar alignment} \cite{Ferreira:2017tvy}.
This means that the direction along the vev alignment is a mass eigenstate which, therefore, couples
to the $WW$ and $ZZ$ pairs just as in the SM.
The other neutral Higgs bosons do not couple to gauge-boson pairs.

The fact that the symmetry group $\Sigma(36)$ is not broken completely by the vacuum configuration means that 
one can classify the physical Higgs bosons according to their conserved charges.
For example, the vev alignment $(1,0,0)$ corresponding to point $B$ preserves
the symmetry group $S_3$ generated by $a$ and $d^2$.
Thus, within each subspace of physical scalar fields (the charged, the light neutral, and the heavy neutral Higgses)
we can construct states which are eigenstates of parity under $d^2$ or which have definite $Z_3$-charges under $a$.
Either of these numbers is conserved. Thus, the lightest pair of states from the second and third doublet
is stable against decay to the SM fields (provided they do not couple to fermions).

\subsection{The scalar sector of $\Sigma(36)$ 3HDM: a summary}

To summarize the above observations, we list here the structural features of the $\Sigma(36)$-symmetric 3HDM scalar sector.
\begin{itemize}
	\item 
	The vev alignment at the global minimum can only be of types \eqref{points-A} to \eqref{points-C}.
	\item
	Spontaneous $CP$ violation is impossible.
	\item
	The model contains automatic scalar alignment, with the SM-like Higgs $h_{SM}$.
	\item
	All charged Higgses are degenerate, and the four non-SM-like neutral Higgs bosons are pair-wise degenerate.
	\item
	The masses of the two pairs of the neutral Higgses are related as $m_{H}^2 = 3 m_h^2$.
	\item
	Since the full symmetry group $\Sigma(36)$ is broken only partially at any of the vev alignments, 
	the lightest non-SM-like Higgs bosons are stable against decay to the SM fields.
\end{itemize}


\section{Alignment preserving soft breaking}\label{section-soft}

The exact discrete symmetry group $\Sigma(36)$ leads to very rigid predictions which could easily 
run in conflict with experiment. It is customary to introduce some flexibility to a model via 
soft breaking terms, which in the case of 3HDM involve up to 9 new free parameters, see Eq.~\eqref{soft}.
The main challenge then is to understand how these soft breaking terms change the structural properties
of the $\Sigma(36)$ scalar sector outlined in the previous section.

Of course, for any specific set of $m_{ij}^2$, one could numerically compute the vevs 
and the properties of the physical scalar bosons. 
But, as we already mentioned in the introduction, the large number of free parameters makes it difficult and impractical
to blindly track down, via a numerical scan, the modifications of the observables in the entire space of soft breaking parameters.
It is even not clear how these numerical results should be presented.
Thus, the real challenge is to comprehend all these dependences,
to construct a clear vision of which parameters govern numerical deviations and which drive structural changes.

In this section, we take the first step towards this vision.
We identify the soft breaking terms which preserve the vev alignments and then study the effects of such terms.
An important consequence is that the automatic scalar alignment with the SM-like Higgs is preserved.
The analytical derivations are corroborated by numerical computation and accompanied with a qualitative discussion. 

\subsection{How to preserve the vev alignment}

Suppose we pick up one of the vev alignments listed in Eqs.~\eqref{points-A} through \eqref{points-C}.
Which terms in $V_{\rm soft}$ can we introduce to keep the alignment intact?

A straightforward way to answer this question for all alignments, one by one, is to write down
the extremum conditions, solve them and deduce the relations among the parameters $m_{ij}^2$
which protect the chosen vev alignment. This method is not very enlightening.
First, it requires direct computations of derivatives for each individual case.
Second, when it leads to certain constraints on the soft breaking parameters $m_{ij}^2$,
it may remain obscure within what range one is allowed to vary them.
Finally, there may arise particular points which may require special treatment.

Instead, we propose here a simple method which leads to a clear picture for any vev alignment.
Furthermore, it can be applied not only to the $\Sigma(36)$ 3HDM, but to any multi-Higgs potential
with the trivial quadratic part, which includes $A_4$, $S_4$ and $\Delta(54)$ 3HDMs.

First, we remind the reader that, when we have a function which depends on the complex variable $z$ and
its conjugate $z^*$, we can differentiate it with respect to $z$ and $z^*$ independently.
Writing $z = x+iy$, we define the antiholomorphic derivative operator as
\begin{equation}
\frac{\partial}{\partial z^*}=\frac{1}{2}\left(\frac{\partial}{\partial x}+i \frac{\partial}{\partial y}\right),\quad
\frac{\partial z}{\partial z^*}=0\,, \quad 
\frac{\partial z^*}{\partial z^*}=1\,.
\end{equation}
The $\Sigma(36)$-invariant potential $V_0 = - m^2 \phi_i^\dagger \phi_i + V_4$ depends on the complex variables 
$\phi_i$ and $\phi_i^\dagger$. Here, the index $i$ can run over six entries: three upper and three lower components of the doublets.
However, since the minimum is neutral, one can suppress the upper components and assume that $\phi_i = \phi_i^0$, $i=1,2,3$.

The extremization condition for the $\Sigma(36)$-symmetric potential is:
\begin{equation}
\frac{\partial V_0}{\partial \phi_i^*} = - m^2 \phi_i + \frac{\partial V_4}{\partial \phi_i^*} = 0\,.
\end{equation}
Therefore, at the extremum point, we have
\begin{equation}
\frac{\partial V_4}{\partial \phi_i^*}\Bigg|_{V_0\ \rm extremum} = m^2 \phi_i\big|_{V_0\ \rm extremum}\,.
\label{relation-at-extr}
\end{equation}
Now, we add the soft breaking terms of Eq.~\eqref{soft}, which we write compactly as
\begin{equation}
V_{\rm soft} = \phi_i^\dagger M_{ij} \phi_j\,,\quad 
M_{ij} = \mmmatrix{m_{11}^2}{m_{12}^2}{(m_{31}^{2})^*}%
{(m_{12}^2)^*}{m_{22}^2}{m_{23}^2}%
{m_{31}^2}{(m_{23}^2)^*}{m_{33}^2}\,,
\end{equation}
with hermitean matrix $M$.
Extremization condition for the full potential $V=V_0 + V_{\rm soft}$ is
\begin{equation}
\frac{\partial V}{\partial \phi_i^*} = M_{ij}\phi_j - m^2 \phi_i + \frac{\partial V_4}{\partial \phi_i^*} = 0\,.\label{extremum-V}
\end{equation}
In general, the extremum point may have shifted with respect to the symmetric case, 
so we cannot use the relation of Eq.~\eqref{relation-at-extr}.
But we now require that the soft breaking terms preserve the vev alignment up to rescaling: 
$v|_{V\ \rm extremum} = \zeta\cdot v|_{V_0\ \rm extremum}$.
Because of this feature and because the quadratic and quartic terms are homogeneous functions with degrees 2 and 4, respectively, 
we can now state that
\begin{equation}
\frac{\partial V_4}{\partial \phi_i^*}\Bigg|_{V\ \rm extremum} = \zeta^2 \cdot m^2 \phi_i\big|_{V\ \rm extremum}\,.
\label{relation-at-extr-2}
\end{equation}
Therefore, at the point of the extremum of $V$ we can simplify Eq.~\eqref{extremum-V} as
\begin{equation}
M_{ij}\phi_j = (1-\zeta^2)m^2 \phi_i\,.
\end{equation}
We conclude that the soft breaking terms preserve a vev alignment of the original symmetric model 
if and only if this vev alignment is an eigenvector of the corresponding matrix $M$. 
This offers us a method of writing down the most general soft breaking terms which preserve any given vev alignment 
of the symmetric model.

\subsection{An example}

Let us illustrate this method with point $C_1$ whose vev alignment is $(1, 1, 1)$.
We want to establish the form of $M$ which would preserve this alignment.
Suppose $\mu_1$, $\mu_2$, $\mu_3$ are the eigenvalues of $M$,
and the complex vectors $\vec n_1$, $\vec n_2$, $\vec n_3$ are the corresponding orthonormal eigenvectors.
Then $M$ can be written as
\begin{equation}
M_{ij} = \mu_1\, n_{1i} n^*_{1j} + \mu_2\, n_{2i} n^*_{2j} + \mu_3\, n_{3i} n^*_{3j}\,.\label{M-via-eigensystem}
\end{equation}
Given one eigenvector, which is already known, 
\begin{equation}
n_1 = \frac{1}{\sqrt{3}}\triplet{1}{1}{1}\,,
\end{equation}
we can select two other eigenvectors in the subspace orthogonal to $\vec n_1$.
Let us define two orthonormal vectors in this subspace, for example, 
\begin{equation}
e_2=\frac{1}{\sqrt{2}}\triplet{0}{1}{-1}\quad \mbox{and} \quad
e_3=\frac{1}{\sqrt{6}}\triplet{-2}{1}{1}\,.\label{e2e3}
\end{equation}
Both $\{\vec n_2,\vec n_3\}$ and $\{\vec e_2,\vec e_3\}$ form a basis.
Therefore, the two eigenvectors $\vec n_2$ and $\vec n_3$ can be obtained from $\vec e_2$ and $\vec e_3$
with an appropriate unitary transformation within this space:
\begin{equation}
\vec n_i = \mathcal{U}_{ij}\vec e_j\,, \quad i,j = 2,3,\quad
\mbox{where} \quad
\mathcal{U} = \begin{pmatrix} \cos\theta & e^{i \xi} \sin\theta \\ 
-e^{-i \xi} \sin\theta & \cos\theta \end{pmatrix}.
\label{e-to-n}
\end{equation}
Notice that multiplying $\vec n_2$ and $\vec n_3$ with additional phase factors does not affect $M$.
Thus, the most general soft breaking terms preserving the vev alignment of point $A$
are described with the matrix $M$ in Eq.~\eqref{M-via-eigensystem} with the following free parameters:
\begin{equation}
\mu_1 = m^2(1-\zeta^2),\quad \mu_2,\quad \mu_3,\quad \theta,\quad \xi\,. 
\end{equation}
If we insist not only on keeping the vev alignment, but also want to preserve the value of $v$, 
we set $\mu_1 = 0$ and are left with 4 parameters, which
we recast in the following more convenient form: 
\begin{equation}
\Sigma = \mu_2 + \mu_3\,,\quad \delta = \mu_2 - \mu_3\,, \quad \theta,\quad \xi\,. \label{4params}
\end{equation}
All these parameters can vary in their full domains of definitions.

The explicit expressions for the matrix $M$ and individual $m_{ij}^2$ soft breaking parameters 
which preserve the vev alignment $C_1$ and the value of $v$ are
\begin{eqnarray}\label{eq:Mij-C}
M_{11} &=& m_{11}^2 = \frac{1}{3} \left(\Sigma - \delta \cos2\theta\right)  \nonumber \\ 
M_{22} &=& m_{22}^2 = \frac{1}{3} \left[\Sigma + \delta \left(\frac{\sqrt{3}}{2} \sin 2\theta \cos \xi + \frac{1}{2} \cos2\theta \right) \right] \nonumber \\
M_{33} &=& m_{22}^2 = \frac{1}{3} \left[\Sigma + \delta \left(-\frac{\sqrt{3}}{2} \sin 2\theta \cos \xi + \frac{1}{2} \cos2\theta \right) \right] \nonumber \\
M_{12} &=& m_{12}^2 = \frac{1}{6} \left[-\Sigma + \delta (-\sqrt{3} \sin2\theta e^{i \xi}  + \cos2\theta )\right]\nonumber  \\
M_{31} &=& m_{31}^2 = \frac{1}{6} \left[-\Sigma + \delta (\sqrt{3} \sin2\theta e^{-i \xi} + \cos2\theta )\right]\nonumber  \\
M_{23} &=& m_{23}^2 = \frac{1}{6} \left[-\Sigma - \delta( i \sqrt{3} \sin 2\theta\sin \xi + 2\cos2\theta)\right]\,.
\end{eqnarray}

In a similar fashion, we parametrize the soft breaking terms which preserve all other
vev alignments of the $\Sigma(36)$ 3HDM, see Appendix~\ref{appendix-all-minima} for the full list.
Here we only remark that, in each case, there exists ambiguity in choosing the basis vectors $\vec e_2$ and $\vec e_3$
with respect to which we parametrize the matrix $M_{ij}$ via angles $\theta$ and $\xi$. 
We resolve this ambiguity by choosing such vectors that the neutral physical Higgs boson masses to be given below
take exactly the same form at all minima.

\subsection{Physical scalars in the softly broken $\Sigma(36)$ 3HDM}

Parametrizing the vev-preserving soft breaking terms as outlined above,
we computed in each case the mass matrices of the physical Higgs bosons.
A remarkable observation is that for all vev alignments 
and with the above choices of the parametrization procedure, 
we could obtain {\em universal} formulas for masses of the physical Higgs bosons,
valid for all the vev alignments of the parent $\Sigma(36)$-symmetric model.
\begin{itemize}
	\item
	The scalar alignment feature is preserved. Indeed, since the vev alignment was the eigenvector
	of the parent model at its minimum and since it is an eigenvector of the matrix of the soft breaking terms,
	it will remain an eigenvector of the Hessian matrix of the softly broken case.
	\item 
	Since we select $\mu_1 = 0$ to preserve not only the alignment but also the value of $v$,
	the mass of the SM-like Higgs boson is unchanged:
	$m^2_{h_{SM}}=  2 (\lambda_1 + \lambda_3)v^2$ for cases $A$ and $A'$ and 
	$m^2_{h_{SM}}=  2 \lambda_1v^2$ for cases $B$ and $C$,
	just as in Eqs.~\eqref{masses-AA} and \eqref{masses-BC}.
	\item
	The non-standard Higgs bosons cease to be mass degenerate.
	For the charged Higgs bosons, we write $m^2_{H^\pm_i} = m^2_{H^\pm_i}\big|_{\Sigma(36)} + \Delta m^2_{H^\pm_i}$, 
	where $m^2_{H^\pm_i}\big|_{\Sigma(36)}$ are their masses in the parent $\Sigma(36)$-symmetric model given by Eqs.~\eqref{masses-AA} and \eqref{masses-BC},
	and observe the following universal corrections:
	\begin{equation}
	\Delta m^2_{H^\pm_1} = \mu_2 = \frac{\Sigma+\delta}{2}, \quad \Delta m^2_{H^\pm_2} = \mu_3=\frac{ \Sigma-\delta}{2}\,.\label{eq:masses1}
	\end{equation}
	The four non-SM-like neutral Higgs bosons have the following masses:
\begin{eqnarray}\label{eq:masses2}
m^2_{h_1} &=& \frac{1}{2} \left(  2 \lvert \lambda_3 \rvert v^2 + \Sigma  - \sqrt{ (\lambda_3 v^2)^2 + \delta^2 + 2x\lvert \lambda_3 \rvert |\delta |v^2}\right), \\
m^2_{h_2} &=& \frac{1}{2} \left(  2 \lvert \lambda_3 \rvert v^2 + \Sigma  - \sqrt{ (\lambda_3 v^2)^2 + \delta^2 - 2x\lvert \lambda_3 \rvert |\delta |v^2}\right), \\
m^2_{H_1} &=& \frac{1}{2} \left(  2 \lvert \lambda_3 \rvert v^2 + \Sigma  + \sqrt{ (\lambda_3 v^2)^2 + \delta^2 - 2x\lvert \lambda_3 \rvert |\delta |v^2}\right), \\
m^2_{H_2} &=& \frac{1}{2} \left(  2 \lvert \lambda_3 \rvert v^2 + \Sigma  + \sqrt{ (\lambda_3 v^2)^2 + \delta^2 + 2x\lvert \lambda_3 \rvert |\delta |v^2}\right)\,,
\label{eq:massesLast}
\end{eqnarray}
	where the quantity $x \in [0,1]$ is
	\begin{equation}
	x = \sqrt{1- (\sin 2\theta\sin \xi)^2}\,.
	\end{equation}
	Here, we write $\lvert \lambda_3 \rvert$ to cover both cases $A$ and $A'$ ($\lambda_3 < 0$) and cases $B$ and $C$ ($\lambda_3 > 0$). 
\end{itemize}
Apart from splitting, the mass patterns demonstrate two remarkable features. 
The first is the unexpected similarity between the cases $A+A'$ and $B+C$. Indeed, cases $A$ and $A'$ are linked by a symmetry of the parent model,
and therefore, one expects that the appropriately parametrized soft terms would lead to the same results for points $A$ and $A'$.
In a similar way, symmetries link cases $B$ and $C$.
However, there is no symmetry of the model which links the vev alignments from points $A$ or $A'$ to points $B$ or $C$.
Indeed, we see that the expressions for the SM-like and charged Higgs masses are not identical.
Whether this intriguing feature can be explained from the ``symmetries of symmetries'' perspective of \cite{Fallbacher:2015rea}
is an open question which deserves a closer look.

The second feature is that the masses depend not on four, but on {\em three} independent soft breaking parameters:
$\Sigma$, $\delta$, and the combination $\sin2\theta\sin\xi$. In the example of point $C$, this combination quantifies 
the imaginary part of $m_{ij}^2$ in Eq.~\eqref{eq:Mij-C}.
This means that, in the 4D space of vev-preserving soft breaking parameters, there exist lines of identical
Higgs spectra. Moving along these lines, one can adjust {\em additional} features of the model, keeping the masses fixed.

It is interesting to observe that, at $\sin2\theta\sin\xi = 1$ leading to $x = 0$, the four neutral Higgses
again combine into two mass-degenerate pairs. The origin of this degeneracy is the special form 
of the soft breaking terms which satisfy $\sin2\theta\sin\xi = 1$.
Such soft breaking terms, in fact, respect several of the symmetries of the vacuum.
Within the same example $C_1$, the soft breaking matrix takes the form
\begin{equation}
M = \frac{\Sigma}{6}\mmmatrix{2}{-1}{-1}{-1}{2}{-1}{-1}{-1}{2} 
+ \frac{\delta\sqrt{3}}{6}\mmmatrix{0}{-i}{i}{i}{0}{-i}{-i}{i}{0}\,,\label{special-M}
\end{equation}
which is invariant under cyclic permutations as well as an exchange of any two doublets followed by a $CP$ transformation.
These residual symmetries form the group $S_3$ and force the neutral scalars to be pairwise degenerate.

The third observation is the remarkable form of neutral mass splittings:
\begin{equation}
m_{h_2}^2 - m_{h_1}^2 = m_{H_2}^2 - m_{H_1}^2\,.\label{equal-splitting}
\end{equation}
It can be viewed as yet another structural feature inherited from the parent symmetric model.

\subsection{Global vs. local minimum}

The potential of the parent $\Sigma(36)$-invariant 3HDM contains six distinct minima which are linked by the broken
symmetry generators and are degenerate.
When we introduce soft breaking terms, we destroy the symmetry, and the six minima are not equivalent anymore.
Therefore, one can wonder if the minimum which one selects to construct the vev-preserving soft breaking terms
represents the global or a local minimum.
The answer turns out surprisingly simple:
the chosen minimum remains the global one if $\mu_2 > 0$ and $\mu_3 > 0$. Additionally, we verified numerically and found that for negative, but small values of either $\mu_2$ or $\mu_3$, the minimum can remain global. The smallness can be quantified relative to the coefficient of the $SU(3)$-invariant quadratic term, $m^2$.

This feature has a simple explanation.
Suppose we select one particular minimum out of the six degenerate minima and add generic soft breaking terms which preserve this minimum.
The depth of the potential at this particular minimum does not change because $\lr{\phi_i^\dagger}M_{ij}\lr{\phi_j} = 0$ (we used here $\mu_1 = 0$).
At all other points, be they extrema or not, the soft breaking terms add $\lr{\phi_i^\dagger}M_{ij}\lr{\phi_j}$ to the potential.
If $\mu_2 > 0$ and $\mu_3 > 0$, this extra contribution is strictly positive everywhere away from the chosen minimum direction.
Therefore, the chosen minimum is automatically the global one.

It is possible to construct examples in which the global minimum is not unique.
For example, one can set $\mu_2 = 0$ and construct such $M$ that the corresponding eigenvector coincides
with another minimum of the $\Sigma(36)$ symmetric model.
In this case, the soft breaking terms will keep unchanged at least two of the previously degenerate minima.

By the same logic, one can also select a small $\mu_2 < 0$ (keeping $\mu_3 > 0$) and select the eigenvector
{\em not} to pass through any other minima. Then, it is possible to find the values of angles $\theta$, $\xi$
which will result in a second minimum at the same depth as the selected one.
By continuity arguments, we see that the selected minimum can remain the global one even if $\mu_2 < 0$.

This analysis is corroborated by a numerical scan over the vev-preserving soft breaking parameter space, which proceeds as follows.
We take a reference $\Sigma(36)$-symmetric parent model by selecting parameters $\lambda_2 > 0$ (required for the minimum to be neutral) and $\lambda_3$
and then expressing $m^2$ and $\lambda_1$ via known $v$ and the SM-like Higgs mass $m_h$.
Once the reference model is fixed, we select a vev alignment and add soft breaking terms $V_{\rm soft}$ 
which preserve the vev alignment selected as well as its magnitude (that is, we set $\mu_1=0$).
We then scan over the soft breaking parameters $\mu_2$, $\mu_3$ in the range from
$-10^6\ \mbox{GeV}^2$ to $10^6\ \mbox{GeV}^2$ and over $\theta, \xi$ in their entire domains. 
At each point, we numerically search for the global minima and, with this information, 
we can determine whether the selected vev alignment stays the global minimum
or becomes a local minimum with a given choice of soft breaking terms.
Also, at each vev alignment, we numerically compute the masses of the physical scalars;
for the minimum we have selected, these masses are found to agree with Eqs.~\eqref{eq:masses1}--\eqref{eq:massesLast}. 

\begin{figure}
	\centering
	\includegraphics[width=0.7\textwidth]{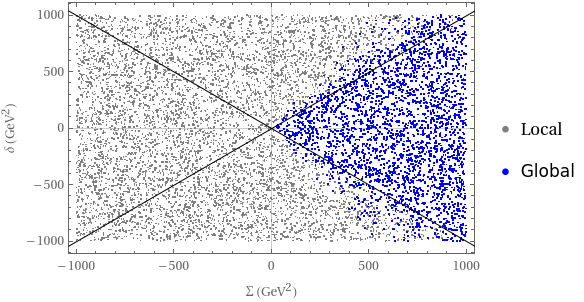}
	\caption{The effect of the vev-preserving alignment parameters of Eq.~\eqref{4params} on the depth of the chosen minimum.
The entire four-parameter scan is projected onto the $(\Sigma, \delta)$ plane within the range 
$-10^3\ \mbox{GeV}^2 \le \Sigma, \delta \le 10^3\ \mbox{GeV}^2$. Blue (gray) dots correspond to soft breaking parameter
sets leading to the global (local) minimum. The black lines are the borders where $\mu_2$ or $\mu_3$ changes sign.}
	\label{Fig:global-local}
\end{figure}

We found that for all points with $\mu_2, \mu_3 > 0$, the chosen minimum remains the global one upon the addition of soft breaking terms.
We also verify that it is the only global minimum of the resulting potential, not a degenerate global minimum.
If $\mu_2$ or $\mu_3$ is significantly negative, the selected minimum unavoidably becomes local.
However we also found many points where one of the two parameters, $\mu_2$ or $\mu_3$, is slightly negative, but the minimum
remains global. Thus, requiring $\mu_2, \mu_3 > 0$ is a sufficient but not necessary assumption for staying in the global minimum.

We illustrate this observation with the scatter plot in Fig.~\ref{Fig:global-local}, where 
we start with the symmetric model with $\lambda_2 = 0.6$, $\lambda_3 = - 0.7$, select a minimum of type A,
perform a four-parameter scan over soft breaking parameters, and project the results onto 
$(\Sigma, \delta)$ plane. We focus here on relatively small values of these two parameters. 
At the black lines, $\mu_2$ or $\mu_3$ changes sign. However the boundary between ``always local'' and ``always global'' parameter
space regions does not coincide with the black lines and is in fact blurred.
A very similar picture is observed for other parameters of the symmetric model.

What we gain from this exercise is the following insight: if one wants to build a softly broken $\Sigma(36)$ 3HDM 
with a vacuum at the brink of absolute tree-level stability,
one should explore the parameter space regions along these lines.

\subsection{Decoupling limits}

Decoupling limit in multi-Higgs models refers to the regime in which additional, non-SM scalars are very heavy,
so that we are left at the electroweak scale with the single Higgs particle whose tree-level properties 
approach the properties if the SM Higgs \cite{Gunion:2002zf}.
Decoupling limit is a weaker statement than the exact decoupling theorem, 
which requires all the effects induced by heavy non-standard
particles to asymptotically disappear in the large mass limit.
It is well known that, in the 2HDM and 3HDM, certain decays of the SM-like Higgs boson 
receive finite corrections from the charged Higgs boson loops
even if their masses are very large \cite{Arhrib:2003vip,Bhattacharyya:2014oka}.
Thus, it is worth scrutinizing the properties of the SM-like Higgs boson in the decoupling limit.

In the 3HDM, we can also define in a similar manner the 2HDM-like limit, when two neutral and a pair of charged Higgses
are heavy and decouple from the remaining relatively light 2HDM-like sector.
Similarly to the distinction between the decoupling theorem and decoupling limit, 
decays of the scalars in this 2HDM-like sector may show deviations with respect to the 2HDM 
which would mimic the mass spectrum of the 3HDM with one generation of very heavy Higgses. 
This comparison would require a dedicated work.

Whether a multi-Higgs model can exhibit the SM decoupling limit depends on its symmetry content.
The recent studies \cite{Nebot:2019qvr,Faro:2020qyp,Carrolo:2021euy} proved that 
a symmetry-constrained multi-Higgs-doublet model allows for the decoupling limit
only when the vev alignment preserves the symmetry group.
In the 3HDMs with Higgs doublets in the 3D irreducible representation of the global symmetry group $G$, 
including the case of $\Sigma(36)$, the vev alignment unavoidably breaks the symmetry group,
which makes the decoupling limit unattainable.
This is also clearly seen by the single quadratic parameter $m^2$ in the $\Sigma(36)$-invariant 3HDM.

The presence of soft breaking terms lead to models without any exact symmetry and, therefore, can display 
the decoupling regime. If both $\mu_2, \mu_3 \gg |\lambda_3|v^2$, 
one can expand the neutral Higgs masses Eqs.~\eqref{eq:masses1}--\eqref{eq:massesLast} as
\begin{equation}
\label{eq:masses_decoupling}
m^2_{h_1, h_2} \approx \mu_2 + |\lambda_3|v^2 \mp \frac{x}{2}|\lambda_3|v^2\,,\quad
m^2_{H_1, H_2} \approx \mu_3 + |\lambda_3|v^2 \mp \frac{x}{2}|\lambda_3|v^2\,.
\end{equation}

One observes the natural scale separation for the two heavy ``multiplets'': 
the squared-masses  of $H_1^\pm$, $h_1$ and $h_2$
stay at the scale $\mu_2$, while $H_2^\pm$, $H_1$ and $H_2$ reside at the scale $\mu_3$.
Within each multiplet, one observes the same mass splitting pattern: 
\begin{equation}
m_{H_1^\pm}^2 - m_{h_1}^2 = \frac{1}{2}v^2\left[\lambda_2 + \lambda_3 f(x)\right]\,,\quad
m_{h_2}^2 - m_{h_1}^2 = x|\lambda_3|v^2\,,\label{pattern}
\end{equation}
with $f(x)= x+1$ for $\lambda_3 > 0$ (points $B$ and $C$) and $f(x)= 2-x$ for $\lambda_3 < 0$ (points $A$ and $A'$),
and exactly the same splitting for $H_2^\pm$, $H_1$, $H_2$.
Notice that since $0\le x \le 1$, the function $f(x)$ lies between 1 and 2 for any choice of the minimum.

Thus, we observe another structural feature of the softly broken model driven by the large symmetry of the parent model: 
in the SM-like decoupling limit, the decoupled sector has a rigid structure of its mass spectrum.

For the 2HDM-like decoupling limit, we assume that $\mu_3$ is large, while $\mu_2$ is of the same order of
magnitude as $\lambda_3 v^2$. The approximate results of Eq.~\eqref{eq:masses_decoupling} driven by large $|\delta|$ remain valid, 
but the mass scales of $h_i$ and $H_i$ are now different. 
The heavy scalars may be dynamically decoupled from the lighter degrees of freedom,
but this does not mean their spectrum can be arbitrary. 
In fact, the heavy scalars display the same pattern of mass splitting shown in Eq.~\eqref{pattern} as the lighter Higgses.
Put simply, decoupling does not imply structural independence of the two sectors.

We also stress that the 2HDM-like model emerging after decoupling of the heavier scalars
is not the general 2HDM but a rather constrained version of it. 
It closely resembles the 2HDM equipped with an approximate $\Z_2$ symmetry
and further constrained by additional relations among parameters. 
Investigation of the phenomenological features of the resulting 2HDM-like model deserves a dedicated study.

\subsection{Decays of non-standard Higgses}

In the parent $\Sigma(36)$ 3HDM, each of the possible minima is still invariant under a subgroup of $\Sigma(36)$.
As a result, the scalar spectrum contains states stabilized against decay by these residual symmetries.
In particular, tree-level trilinear couplings of such states to $h_{SM}$ pairs are all vanishing.
However, the vev alignment preserving soft breaking terms, in general, remove all the symmetries from the model.
As a result, the Higgses which were previously stabilized by residual symmetries are not protected anymore
and can decay to the SM-like Higgses and further to the SM fields. 

To understand how these decays proceed, suppose that $h_1$ is the lightest non-SM-like scalar.
In the parent symmetric model, we had very few trilinear couplings involving $h_1$.
With the soft breaking terms, one adds a few more terms, but the interaction vertices $h_1 h_{SM}h_{SM}$ 
and $h_1 h_{SM}h_{SM}h_{SM}$ which could generate tree-level decays of $h_1$ are still absent.
Thus, there is no tree-level path to the decay of $h_1$.

\begin{figure}[!h]
	\centering
	\includegraphics[width=0.8\textwidth]{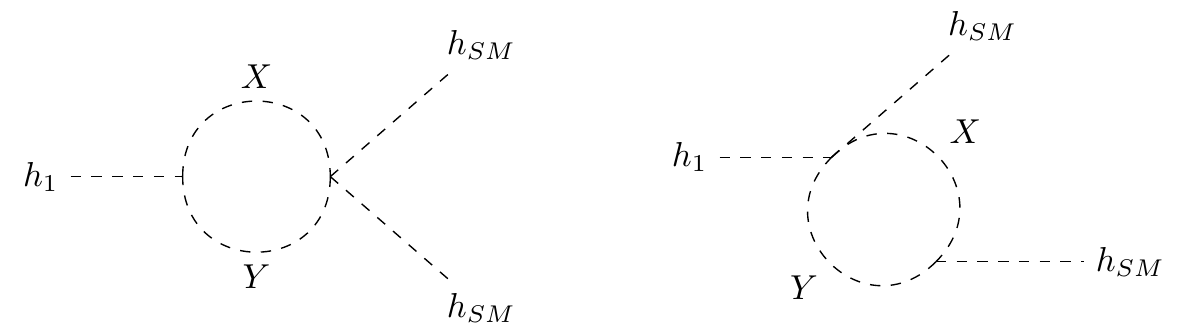}
\caption{Scalar loop diagrams inducing $h_1 \to h_{SM}h_{SM}$ decays in the softly broken $\Sigma(36)$ 3HDM, where $X, Y$ denote any scalar field.}
\label{Fig:diagrams}
\end{figure}

Next, we checked scalar combinations which could lead to one-loop decays through the diagrams
shown in Fig.~\ref{Fig:diagrams}.
We found that there exist matching trilinear and quartic couplings
($h_1 X Y$ and $X Y h_{SM} h_{SM}$ for topology 1, $h_1 h_{SM} X Y$ and $X Y h_{SM}$ for topology 2)
which share the same pairs of scalars $XY$. 
These matching pairs appear only with soft breaking terms; they were absent in the symmetric model. 

These diagrams induce decay of $h_1$, which may be suppressed due to a number of reasons
(loop factors, small couplings, subthreshold effects for $m_{h_1} < 2 m_{h_{SM}}$).
If this suppression is significant, it may lead to displaced vertex signals which could be seen at colliders.
Calculation of these decays should be an important part of a detailed phenomenological study of the 
softly broken $\Sigma(36)$ 3HDM.

\section{Discussion and conclusions}

Historically, multi-Higgs-doublet models with large symmetry groups triggered interest 
thanks to the opportunities they offered to link hierarchical quark masses and mixing patterns 
as well as the amount of $CP$ violation with symmetry group properties.
It turned out, however, that large exact discrete symmetry groups are too restrictive
and run into conflict with quark properties \cite{Leurer:1992wg,GonzalezFelipe:2014mcf}.
This obstacle can be avoided if the large symmetry group is softly broken by quadratic terms in the potential.
However, the large number of new free parameters associated with the general soft breaking terms come 
makes a straightforward analysis of their consequences --- and even the presentations of the results --- rather cumbersome.
One needs additional methods capable of indicating which directions in the soft breaking parameter space
are linked to which kind of phenomenological signals.

In this paper, we began developing such methods. Relying on the fact that multi-Higgs models with large discrete symmetry groups
lead to very specific vev alignments, we asked which soft breaking terms could preserve a chosen alignment and
found a general constructive answer through an eigenvector-based procedure.

We illustrated this procedure with the 3HDM example based on the softly broken symmetry group $\Sigma(36)$.
Out of the nine soft breaking parameters, we identified five which preserve the vev alignment
and four which break it. Focusing on the vev-alignment preserving terms,
we investigated scalar alignment, physical Higgs masses and their relations, the global vs. local minimum distinction,
stable vs. unstable scalars, existence of the SM-like and 2HDM-like decoupling limits.
Remarkably, although the softly broken model does not possess any exact symmetry,
we found that is still possesses several structural properties inherited from the parent $\Sigma(36)$-symmetric 3HDM.
They included scalar alignment, certain relations among Higgs masses, and peculiar form of decoupling to a 2HDM-like model
(that is, decoupling does not imply complete independence).

The vision which emerges from this study will guide further detailed phenomenological studies of softly broken symmetry models.
If one asks for specific signatures from softly broken symmetries, this procedure will indicate
which parameters must be taken into account and which are inessential.
More in-depth phenomenological investigations of the resulting benchmark models will be a subject of future works.

We also believe that the ``symmetries of symmetries'' formalism to the structural properties 
of the symmetric scalar potentials developed in \cite{Fallbacher:2015rea} can provide additional insights into this problem. 

\section*{Acknowledgments}
We thank Andreas Trautner and the referee for their useful comments. 
We acknowledge funding from the Portuguese
\textit{Fun\-da\-\c{c}\~{a}o para a Ci\^{e}ncia e a Tecnologia} (FCT) through the FCT Investigator 
contracts IF/00816/2015 and through the projects UID/FIS/00777/2013, UID/FIS/00777/2019, 
CERN/FIS-PAR/0004/2019, and PTDC/FIS-PAR/29436/2017,
which are partially funded through POCTI (FEDER), COMPETE, QREN, and the EU.
The work of ML is also funded by
Funda\c{c}\~{a}o para a Ci\^{e}ncia e Tecnologia-FCT through Grant
No.PD/BD/150488/2019, in the framework of the Doctoral Programme IDPASC-PT.
We also acknowledge the support from National Science Center, Poland, via the project Harmonia (UMO-2015/18/M/ST2/00518).

\appendix

\section{$\Sigma(36)$ vs. $\Delta(54)$ 3HDM}\label{appendix-Delta54}

To avoid confusion, let us clarify the definition of the group $\Sigma(36)$.
If one understands the generators $a$, $b$, and $d$ given in Eq. ~\eqref{Sigma36-generators}
as transformations, from $SU(3)$, then the group generated by them is
\begin{equation}
\Sigma(36\varphi) \simeq \Delta(27)\rtimes \Z_4\,,
\end{equation}
which has order $|\Sigma(36\varphi)| = 108$. However, $SU(3)$ contains its center, the group
$\Z_3$ generated by $\omega(1,1,1)$, which belongs to the global group hypercharge transformation group.
Factoring $SU(3)$ by its center brings us to $PSU(3) \simeq SU(3)/\Z_3$.
The group $\Sigma(36) \simeq \Sigma(36\varphi)/\Z_3$ of order 36 is understood as the subgroup of $PSU(3)$.

When defining a group in $PSU(3)$, one can still write a generator $g$ as a unitary $3\times 3$
matrix, which is understood as a representative point of the entire coset $g\cdot \Z_3$.
Thus, one can still use the generators $a$, $b$, and $d$ as in Eq.~\eqref{Sigma36-generators},
provided one considers their relations up to any possible transformation from the center.
It is in this sense that we say that the generators $a$ and $b$ commute: their commutator
$aba^{-1}b^{-1}$ produces an element from the center of $SU(3)$, which becomes an identity element of $PSU(3)$.
For more discussion of these subtle distinctions, see \cite{Ivanov:2011ae,Ivanov:2012fp}.

We remark that the traditional notation of symmetry groups in the scalar sector of 3HDM inadvertently confuses 
the two spaces. That is, when one defines the $A_4$ 3HDM, the group $A_4$ is understood as a subgroup of $PSU(3)$,
while when one speaks of $\Delta(27)$ 3HDM, one uses $\Delta(27)$ which is a subgroup of $SU(3)$.

The symmetry group $\Sigma(36)$ is twice larger than the more familiar group $\Delta(54)$ 
(which is, in fact, just $(\Z_3\times\Z_3)\rtimes \Z_2$ inside $PSU(3)$).
One would obtain the $\Delta(54)$-symmetric 3HDM, if one required invariance under the generator $d^2$, not $d$ itself.
$\Delta(54)$ allows for additional terms in the potential which are absent in Eq.~\eqref{Vexact}.

In the $CP$-violating version of $\Delta(54)$ 3HDM, 
the three points $A$ would not be linked with $A'$ by a symmetry transformation.
The same would apply to the points $B$ and $C$. Thus, in $CP$-violating $\Delta(54)$ 3HDM,
depending on the values of the parameters, the minimum could be at 
$A$, $A'$, $B$, or $C$.
For the $CP$-conserving $\Delta(54)$, points $A$ and $A'$ become related by a (generalized) $CP$ symmetry transformation,
so the minimum can be either at points $A + A'$ or $B$ or $C$. 
With the enhanced family symmetry $\Sigma(36)$ points $B$ and $C$ become equivalent, too.

\section{Alignment preserving soft breaking terms for all the minima}\label{appendix-all-minima}

For completeness, we list here the explicit expressions for eigenvectors and the parametrization
of the soft breaking terms $M_{ij}$ for all the minima of the $\Sigma(36)$ symmetric model.

We begin the case considered in the main text and then use it to build all other cases of type $C$, $A$ and $A'$:
\begin{itemize}
\item
For point $C_1$ with the alignment $(1,1,1)$ we use
\begin{equation}
n_1 = \frac{1}{\sqrt{3}}\triplet{1}{1}{1}\,,\quad
e_2=\frac{1}{\sqrt{2}}\triplet{0}{1}{-1}\,,\quad 
e_3=\frac{1}{\sqrt{6}}\triplet{-2}{1}{1}\,,\label{n1e2e3-C1}
\end{equation}
and construct the vectors $\vec n_2$, $\vec n_3$ using the matrix in Eq.~\eqref{e-to-n}.
The hermitean matrix $M_{ij}$ has the following elements:
\begin{eqnarray}\label{Mij-C1}
M_{11} &=& m_{11}^2 = \frac{1}{3} \left(\Sigma - \delta \cos2\theta\right)  \nonumber \\ 
M_{22} &=& m_{22}^2 = \frac{1}{3} \left[\Sigma + \delta \left(\frac{\sqrt{3}}{2} \sin 2\theta \cos \xi + \frac{1}{2} \cos2\theta \right) \right] \nonumber \\
M_{33} &=& m_{22}^2 = \frac{1}{3} \left[\Sigma + \delta \left(-\frac{\sqrt{3}}{2} \sin 2\theta \cos \xi + \frac{1}{2} \cos2\theta \right) \right] \nonumber \\
M_{12} &=& m_{12}^2 = \frac{1}{6} \left[-\Sigma + \delta (-\sqrt{3} \sin2\theta e^{i \xi}  + \cos2\theta )\right]\nonumber  \\
M_{31} &=& m_{31}^2 = \frac{1}{6} \left[-\Sigma + \delta (\sqrt{3} \sin2\theta e^{-i \xi} + \cos2\theta )\right]\nonumber  \\
M_{23} &=& m_{23}^2 = \frac{1}{6} \left[-\Sigma - \delta( i \sqrt{3} \sin 2\theta\sin \xi + 2\cos2\theta)\right]\,.
\end{eqnarray}
\item
For point $C_2$ with the alignment $(1,\omega,\omega^2)$ we use:
\begin{equation}
n_1 = \frac{1}{\sqrt{3}}\triplet{1}{\omega}{\omega^2}\,,\quad
e_2=\frac{1}{\sqrt{2}}\triplet{0}{\omega}{-\omega^2}\,.\quad 
e_3=\frac{1}{\sqrt{6}}\triplet{-2}{\omega}{\omega^2}\,.\label{n1e2e3-C2}
\end{equation}
The hermitean matrix $M_{ij}$ can now be easily expressed as 
\begin{equation}
M_{ij}\Big|_{C_2} = \mmmatrix{\dots}{\omega^2 \dots}{\omega\dots}{\omega\dots}{\dots}{\omega^2 \dots}{\omega^2 \dots}{\omega\dots}{\dots}\,,\label{Mij-C2}
\end{equation}
where dots indicate the corresponding element of the matrix $M_{ij}$ at point $C_1$ given in Eq.~\eqref{Mij-C1}.
\item
For point $C_3$ with the alignment $(1,\omega^2,\omega)$ we use:
\begin{equation}
n_1 = \frac{1}{\sqrt{3}}\triplet{1}{\omega^2}{\omega}\,,\quad
e_2=\frac{1}{\sqrt{2}}\triplet{0}{\omega^2}{-\omega}\,.\quad 
e_3=\frac{1}{\sqrt{6}}\triplet{-2}{\omega^2}{\omega}\,.\label{n1e2e3-C3}
\end{equation}
The elements of the hermitean matrix $M_{ij}$ are now 
\begin{equation}
M_{ij}\Big|_{C_3} = \mmmatrix{\dots}{\omega \dots}{\omega^2\dots}{\omega^2\dots}{\dots}{\omega \dots}{\omega \dots}{\omega^2\dots}{\dots}\,,\label{Mij-C3}
\end{equation}
	\item
For point $A_1$ with the alignment $(\omega,1,1)$ we use:
\begin{equation}
n_1 = \frac{1}{\sqrt{3}}\triplet{\omega}{1}{1}\,,\quad
e_2=\frac{1}{\sqrt{2}}\triplet{0}{1}{-1}\,.\quad 
e_3=\frac{1}{\sqrt{6}}\triplet{-2\omega}{1}{1}\,.\label{n1e2e3-A1}
\end{equation}
The hermitean matrix $M_{ij}$ is now  
\begin{equation}
M_{ij}\Big|_{A_1} = \mmmatrix{\dots}{\omega \dots}{\omega\dots}{\omega^2\dots}{\dots}{\dots}{\omega^2 \dots}{\dots}{\dots}\,,\label{Mij-A1}
\end{equation}
	\item
For point $A_2$ with the alignment $(1,\omega,1)$ we use:
\begin{equation}
n_1 = \frac{1}{\sqrt{3}}\triplet{1}{\omega}{1}\,,\quad
e_2=\frac{1}{\sqrt{2}}\triplet{0}{\omega}{-1}\,.\quad 
e_3=\frac{1}{\sqrt{6}}\triplet{-2}{\omega}{1}\,.\label{n1e2e3-A2}
\end{equation}
The hermitean matrix $M_{ij}$ is now  
\begin{equation}
M_{ij}\Big|_{A_2} = \mmmatrix{\dots}{\omega^2\dots}{\dots}{\omega\dots}{\dots}{\omega\dots}{\dots}{\omega^2\dots}{\dots}\,,\label{Mij-A2}
\end{equation}
	\item
For point $A_3$ with the alignment $(1,1,\omega)$ we use:
\begin{equation}
n_1 = \frac{1}{\sqrt{3}}\triplet{1}{1}{\omega}\,,\quad
e_2=\frac{1}{\sqrt{2}}\triplet{0}{1}{-\omega}\,.\quad 
e_3=\frac{1}{\sqrt{6}}\triplet{-2}{1}{\omega}\,.\label{n1e2e3-A3}
\end{equation}
The hermitean matrix $M_{ij}$ is now  
\begin{equation}
M_{ij}\Big|_{A_3} = \mmmatrix{\dots}{\dots}{\omega^2\dots}{\dots}{\dots}{\omega^2\dots}{\omega\dots}{\omega\dots}{\dots}\,,\label{Mij-A3}
\end{equation}
	\item
For points $A'_1$, $A'_2$, $A'_3$, we obtain the relevant expressions by performing complex conjugation (not hermitean conjugation!) 
of the corresponding expressions for points $A_1$, $A_2$, $A_3$.
\end{itemize}

Finally, for points of type $B$ we use a slightly different choice of basis eigenvectors.
\begin{itemize}
	\item
For point $B_1$ with alignment $(1,0,0)$ we use
\begin{equation}
n_1 = \triplet{1}{0}{0}\,,\quad
e_2 =\frac{1}{\sqrt{2}} \triplet{0}{1}{i}\,,\quad
e_3 = \frac{1}{\sqrt{2}}\triplet{0}{i}{1}\,.\label{n1e2e3-B1}
\end{equation}
The resulting matrix $M_{ij}$ has the following elements:
\begin{eqnarray}\label{eq:Mij-B1}
&& M_{11} = M_{12} = M_{13} = 0 \nonumber \\
&&M_{22} = \frac{1}{2} \left( \Sigma - \delta \sin2\theta \sin\xi\right) \nonumber\\
&&M_{33} = \frac{1}{2} \left( \Sigma + \delta \sin2\theta \sin\xi\right) \nonumber\\
&&M_{23} = \frac{1}{2} \delta \left(\sin{2\theta} \cos{\xi} - i \cos{2\theta} \right)\,.
\end{eqnarray}
The motivation for the choice \eqref{n1e2e3-A3} is the following.
In all previous cases, by setting $\sin2\theta\sin\xi = 1$, we obtain soft breaking terms which
respect several symmetries of the vacuum which, in turn, leads to pairwise mass degenerate neutral scalars,
see the discussion around Eq.~\eqref{special-M}.
We want to achieve the same feature for points $B$. This can be done if $M$ is diagonal
(the preserved symmetries being the generator $a$ in Eq.~\eqref{Sigma36-generators} and the ordinary $CP$).
The choice of Eq.~\eqref{n1e2e3-B1} is exactly the one which produces diagonal $M$ for $\sin2\theta\sin\xi = 1$.
	\item
For point $B_2$ with alignment $(0,1,0)$ we use
\begin{equation}
n_1 = \triplet{0}{1}{0}\,,\quad
e_2 =\frac{1}{\sqrt{2}} \triplet{i}{0}{1}\,,\quad
e_3 = \frac{1}{\sqrt{2}}\triplet{1}{0}{i}\,.
\end{equation}
The resulting matrix $M_{ij}$ has the following elements:
\begin{eqnarray}\label{eq:Mij-B2}
&& M_{21} = M_{22} = M_{23} = 0 \nonumber \\
&&M_{33} = \frac{1}{2} \left( \Sigma - \delta \sin2\theta \sin\xi\right) \nonumber\\
&&M_{11} = \frac{1}{2} \left( \Sigma + \delta \sin2\theta \sin\xi\right) \nonumber\\
&&M_{31} = \frac{1}{2} \delta \left(\sin{2\theta} \cos{\xi} - i \cos{2\theta} \right)\,.
\end{eqnarray}
	\item
For point $B_3$ with alignment $(0,0,1)$ we use
\begin{equation}
n_1 = \triplet{0}{0}{1}\,,\quad
e_2 =\frac{1}{\sqrt{2}} \triplet{1}{i}{0}\,,\quad
e_3 = \frac{1}{\sqrt{2}}\triplet{i}{1}{0}\,.
\end{equation}
The resulting matrix $M_{ij}$ has the following elements:
\begin{eqnarray}\label{eq:Mij-B}
&& M_{31} = M_{32} = M_{33} = 0 \nonumber \\
&&M_{11} = \frac{1}{2} \left( \Sigma - \delta \sin2\theta \sin\xi\right) \nonumber\\
&&M_{22} = \frac{1}{2} \left( \Sigma + \delta \sin2\theta \sin\xi\right) \nonumber\\
&&M_{12} = \frac{1}{2} \delta \left(\sin{2\theta} \cos{\xi} - i \cos{2\theta} \right)\,.
\end{eqnarray}
\end{itemize}


\begin{thebibliography}{99}
\bibitem{Branco:2011iw}
G.~C.~Branco, P.~M.~Ferreira, L.~Lavoura, M.~N.~Rebelo, M.~Sher and J.~P.~Silva,
Phys. Rept. \textbf{516}, 1-102 (2012)
[arXiv:1106.0034 [hep-ph]].

\bibitem{Kanemura:2014bqa}
S.~Kanemura, K.~Tsumura, K.~Yagyu and H.~Yokoya,
Phys. Rev. D \textbf{90}, 075001 (2014)
[arXiv:1406.3294 [hep-ph]].

\bibitem{Ivanov:2017dad}
I.~P.~Ivanov,
Prog. Part. Nucl. Phys. \textbf{95}, 160-208 (2017)
[arXiv:1702.03776 [hep-ph]].

\bibitem{Arcadi:2019lka}
G.~Arcadi, A.~Djouadi and M.~Raidal,
Phys. Rept. \textbf{842}, 1-180 (2020)
[arXiv:1903.03616 [hep-ph]].

\bibitem{Weinberg:1976hu}
S.~Weinberg,
Phys. Rev. Lett. \textbf{37}, 657 (1976).

\bibitem{Leurer:1992wg}
M.~Leurer, Y.~Nir and N.~Seiberg,
Nucl. Phys. B \textbf{398}, 319-342 (1993)
[arXiv:hep-ph/9212278 [hep-ph]].

\bibitem{GonzalezFelipe:2014mcf}
R.~Gonz\'alez Felipe, I.~P.~Ivanov, C.~C.~Nishi, H.~Ser\^odio and J.~P.~Silva,
Eur. Phys. J. C \textbf{74}, no.7, 2953 (2014)
[arXiv:1401.5807 [hep-ph]].

\bibitem{Ivanov:2012fp}
I.~P.~Ivanov and E.~Vdovin,
Eur. Phys. J. C \textbf{73}, no.2, 2309 (2013)
[arXiv:1210.6553 [hep-ph]].

\bibitem{Darvishi:2019dbh}
N.~Darvishi and A.~Pilaftsis,
Phys. Rev. D \textbf{101}, no.9, 095008 (2020)
[arXiv:1912.00887 [hep-ph]].

\bibitem{Darvishi:2021txa}
N.~Darvishi, M.~R.~Masouminia and A.~Pilaftsis,
[arXiv:2106.03159 [hep-ph]].

\bibitem{Nishi:2006tg}
C.~C.~Nishi,
Phys. Rev. D \textbf{74}, 036003 (2006)
[erratum: Phys. Rev. D \textbf{76}, 119901 (2007)]
[arXiv:hep-ph/0605153 [hep-ph]].

\bibitem{deMedeirosVarzielas:2017ote}
I.~de Medeiros Varzielas, S.~F.~King, C.~Luhn and T.~Neder,
JHEP \textbf{11}, 136 (2017)
[arXiv:1706.07606 [hep-ph]].

\bibitem{Ivanov:2018ime}
I.~P.~Ivanov, C.~C.~Nishi, J.~P.~Silva and A.~Trautner,
Phys. Rev. D \textbf{99}, no.1, 015039 (2019)
[arXiv:1810.13396 [hep-ph]].

\bibitem{Ivanov:2019kyh}
I.~P.~Ivanov, C.~C.~Nishi and A.~Trautner,
Eur. Phys. J. C \textbf{79}, no.4, 315 (2019)
[arXiv:1901.11472 [hep-ph]].

\bibitem{deMedeirosVarzielas:2019rrp}
I.~de Medeiros Varzielas and I.~P.~Ivanov,
Phys. Rev. D \textbf{100}, no.1, 015008 (2019)
[arXiv:1903.11110 [hep-ph]].

\bibitem{Ivanov:2014doa}
I.~P.~Ivanov and C.~C.~Nishi,
JHEP \textbf{01}, 021 (2015)
[arXiv:1410.6139 [hep-ph]].

\bibitem{deMedeirosVarzielas:2017glw}
I.~de Medeiros Varzielas, S.~F.~King, C.~Luhn and T.~Neder,
Phys. Lett. B \textbf{775}, 303-310 (2017)
[arXiv:1704.06322 [hep-ph]].

\bibitem{Segre:1978ji}
G.~Segre and H.~A.~Weldon,
Phys. Lett. B \textbf{83}, 351-354 (1979).

\bibitem{Grimus:2010ak}
W.~Grimus and P.~O.~Ludl,
J. Phys. A \textbf{43}, 445209 (2010)
[arXiv:1006.0098 [hep-ph]].

\bibitem{Merle:2011vy}
A.~Merle and R.~Zwicky,
JHEP \textbf{02}, 128 (2012)
[arXiv:1110.4891 [hep-ph]].

\bibitem{Hagedorn:2013nra}
C.~Hagedorn, A.~Meroni and L.~Vitale,
J. Phys. A \textbf{47}, 055201 (2014)
[arXiv:1307.5308 [hep-ph]].

\bibitem{Rong:2016cpk}
S.~J.~Rong,
Phys. Rev. D \textbf{95}, no.7, 076014 (2017)
[arXiv:1604.08482 [hep-ph]].

\bibitem{Ivanov:2011ae}
I.~P.~Ivanov, V.~Keus and E.~Vdovin,
J. Phys. A \textbf{45}, 215201 (2012)
[arXiv:1112.1660 [math-ph]].

\bibitem{Branco:1983tn}
G.~C.~Branco, J.~M.~Gerard and W.~Grimus,
Phys. Lett. B \textbf{136}, 383-386 (1984).

\bibitem{deMedeirosVarzielas:2011zw}
I.~de Medeiros Varzielas and D.~Emmanuel-Costa,
Phys. Rev. D \textbf{84}, 117901 (2011)
[arXiv:1106.5477 [hep-ph]].

\bibitem{Varzielas:2012nn}
I.~de Medeiros Varzielas, D.~Emmanuel-Costa and P.~Leser,
Phys. Lett. B \textbf{716}, 193-196 (2012)
[arXiv:1204.3633 [hep-ph]].

\bibitem{Ivanov:2013nla}
I.~P.~Ivanov and L.~Lavoura,
Eur. Phys. J. C \textbf{73}, no.4, 2416 (2013)
[arXiv:1302.3656 [hep-ph]].

\bibitem{Bhattacharyya:2012pi}
G.~Bhattacharyya, I.~de Medeiros Varzielas and P.~Leser,
Phys. Rev. Lett. \textbf{109}, 241603 (2012)
[arXiv:1210.0545 [hep-ph]].

\bibitem{Varzielas:2013sla}
I.~de Medeiros Varzielas and D.~Pidt,
J. Phys. G \textbf{41}, 025004 (2014)
[arXiv:1307.0711 [hep-ph]].

\bibitem{Varzielas:2013eta}
I.~de Medeiros Varzielas and D.~Pidt,
JHEP \textbf{11}, 206 (2013)
[arXiv:1307.6545 [hep-ph]].

\bibitem{Fallbacher:2015rea}
M.~Fallbacher and A.~Trautner,
Nucl. Phys. B \textbf{894}, 136-160 (2015)
[arXiv:1502.01829 [hep-ph]].

\bibitem{Ferreira:2017tvy}
P.~M.~Ferreira, I.~P.~Ivanov, E.~Jim\'enez, R.~Pasechnik and H.~Ser\^odio,
JHEP \textbf{01}, 065 (2018)
[arXiv:1711.02042 [hep-ph]].

\bibitem{Gunion:2002zf}
J.~F.~Gunion and H.~E.~Haber,
Phys. Rev. D \textbf{67}, 075019 (2003)
[arXiv:hep-ph/0207010 [hep-ph]].

\bibitem{Arhrib:2003vip}
A.~Arhrib, M.~Capdequi Peyranere, W.~Hollik and S.~Penaranda,
Phys. Lett. B \textbf{579}, 361-370 (2004)
[arXiv:hep-ph/0307391 [hep-ph]].

\bibitem{Bhattacharyya:2014oka}
G.~Bhattacharyya and D.~Das,
Phys. Rev. D \textbf{91}, 015005 (2015)
[arXiv:1408.6133 [hep-ph]].

\bibitem{Nebot:2019qvr}
M.~Nebot,
Phys. Rev. D \textbf{102}, no.11, 115002 (2020)
[arXiv:1911.02266 [hep-ph]].

\bibitem{Faro:2020qyp}
F.~Faro, J.~C.~Romao and J.~P.~Silva,
Eur. Phys. J. C \textbf{80}, no.7, 635 (2020)
[arXiv:2002.10518 [hep-ph]].

\bibitem{Carrolo:2021euy}
S.~Carrolo, J.~C.~Rom\~ao, J.~P.~Silva and F.~Vaz\~ao,
Phys. Rev. D \textbf{103}, no.7, 075026 (2021)
[arXiv:2102.11303 [hep-ph]].
  
\end{thebibliography}
\end{document}